\documentclass[twocolumn,english,aps,prl,showpacs,floatfix,superscriptaddress]{revtex4}
\usepackage[T1]{fontenc}
\usepackage[latin9]{inputenc}
\usepackage{amsmath}
\usepackage{amssymb}
\usepackage{graphicx}

\usepackage[normalem]{ulem}
\usepackage{color}
\definecolor{darkerblue}{rgb}{0,0,0.75}
\definecolor{darkerred}{rgb}{0.8,0,0}

\usepackage[draft]{hyperref}

\makeatletter
\@ifundefined{textcolor}{}
{%
 \definecolor{BLACK}{gray}{0}
 \definecolor{WHITE}{gray}{1}
 \definecolor{RED}{rgb}{1,0,0}
 \definecolor{GREEN}{rgb}{0,1,0}
 \definecolor{BLUE}{rgb}{0,0,1}
 \definecolor{CYAN}{cmyk}{1,0,0,0}
 \definecolor{MAGENTA}{cmyk}{0,1,0,0}
 \definecolor{YELLOW}{cmyk}{0,0,1,0}
}

\usepackage{siunitx}

\@ifundefined{definecolor}{\usepackage{color}}{}

\newcommand{\er}{{\bf r}}

\usepackage{babel}

\makeatother

\usepackage{babel}
\begin{document}
\flushbottom

\title{Josephson vortices in a long Josephson junction formed by phase twist in a polariton superfluid}

\author{Davide~Caputo}
\affiliation{CNR NANOTEC---Institute of Nanotechnology, Via Monteroni, 73100 Lecce, Italy}
\affiliation{University of Salento, Via Arnesano, 73100 Lecce, Italy}

\author{Nataliya~Bobrovska}
\affiliation{Institute of Physics, Polish Academy of Sciences, 02-668 Warszawa, Poland}

\author{Dario~Ballarini}
\affiliation{CNR NANOTEC---Institute of Nanotechnology, Via Monteroni, 73100 Lecce, Italy}

\author{Michal~Matuszewski}
\affiliation{Institute of Physics, Polish Academy of Sciences, 02-668 Warszawa, Poland}

\author{Milena~De~Giorgi}
\affiliation{CNR NANOTEC---Institute of Nanotechnology, Via Monteroni, 73100 Lecce, Italy}

\author{Lorenzo~Dominici}
\affiliation{CNR NANOTEC---Institute of Nanotechnology, Via Monteroni, 73100 Lecce, Italy}

\author{Kenneth~West}
\affiliation{PRISM, Princeton Institute for the Science and Technology of Materials, Princeton Unviversity, Princeton, NJ 08540}

\author{Loren~N.~Pfeiffer}
\affiliation{PRISM, Princeton Institute for the Science and Technology of Materials, Princeton Unviversity, Princeton, NJ 08540}

\author{Giuseppe~Gigli}
\affiliation{CNR NANOTEC---Institute of Nanotechnology, Via Monteroni, 73100 Lecce, Italy}
\affiliation{University of Salento, Via Arnesano, 73100 Lecce, Italy}

\author{Daniele~Sanvitto}
\affiliation{CNR NANOTEC---Institute of Nanotechnology, Via Monteroni, 73100 Lecce, Italy}
\affiliation{INFN, Sez. Lecce, 73100 Lecce, Italy}

\maketitle

{\bf Quantum fluids of light are an emerging platform for energy efficient signal processing, ultra-sensitive interferometry and quantum simulators at elevated temperatures.
Here we demonstrate the optical control of the topological excitations induced in a large polariton condensate, realising the bosonic analog of a long Josephson junction and reporting the first observation of bosonic Josephson vortices. When a phase difference is imposed at the boundaries of the condensate, two extended regions become separated by a sharp $\pi$-slippage of the phase and a solitonic depletion of the density, forming an insulating barrier with a suppressed order parameter. The superfluid behavior, that is a smooth phase gradient across the system instead of the sharp phase jump, is recovered at higher polariton densities and it is mediated by the nucleation of Josephson vortices within the barrier. Our results contribute to the understanding of dissipation and stability of elementary excitations in macroscopic quantum systems.
}

\section*{Introduction}
Bose-Einstein condensation and the closely related phenomena of superfluidity and superconductivity are the most striking manifestations of macroscopic quantum physics. 
The quest for practical applications stimulated the search for new systems in which superconductivity or superfluidity could be exploited in realistic devices. In the last decade, the development of new materials and heterostructures led to the emergence of quantum fluids of light~\cite{QuantumFluids, Sanvitto_PolaritonDevices}. Typically, they are realized in semiconductor microcavities, which allow for the existence of peculiar states called exciton-polaritons~\cite{KasprzakBEC}. 
In these systems, light-matter interaction is larger than dissipation and the elementary excitations cannot be described by the bare exciton and photon modes. Instead, exciton-polaritons are the new eigenstates, resulting from the coherent coupling of excitons and photons. Exciton-polaritons (hereafter, polaritons) possess extremely low effective mass ($10^{-5}\times m_{0}$, with $m_{0}$ the electron mass) and bosonic statistics, providing a suited system to observe condensation in a single state even at room temperature and in a solid state environment. Moreover, their peculiar light-matter composition allows the investigation of collective density and phase excitations with relatively simple optical setups, with the advantage of the direct measurement of the velocity field from the phase gradient~\cite{Nardin2011}. 
Recently, superfluidity has been achieved at room temperature in organic semiconductor structures~\cite{LerarioRTSuperfluidity}, and analogs of short Josephson junction have been demonstrated in inorganic microcavities~\cite{Deveaud_JJ,Bloch_JJ}.

More generally, the behavior of a complex order parameter in proximity of a junction and the associated formation of phase slips and vortices have been at the center of intensive research, not only in superconducting systems, but also in superfluid helium and Bose-Einstein condensates of ultracold atoms~\cite{Sukhatme2001, Tanzi2016,Abad2015}.  
In contrast to the short (point-like) Josephson junctions, the so-called long Josephson junctions (LJJs) are characterized by an interface that is extended beyond the Josephson penetration depth at least in one dimension~\cite{Barone_book}.  
One of the most interesting phenomena occurring in LJJs are Josephson vortices, which, in contrast to Abrikosov or Pearl vortices, are localized within the barrier and characterized by opposite transverse supercurrents in the two superconductors~\cite{Kaurov2005, Roditchev2015,Yoshizawa_ImagingJV}. 
With respect to the charged case, a bosonic Josephson vortex is characterised by the absence of a magnetic flux and its description involves necessarily both the phase and the amplitude of the wavefunction~\cite{Kaurov2006, Abad2011}.

Here, we observe a LJJ in a system composed of two regions of exciton-polariton condensate with macroscopic quantum phases controlled by additional external lasers. The junction interface extends over several tens of microns and forms in response to a twist of the phase of the condensate. 
The twist results in the creation of a dark soliton-like coherent structure, which plays the role of an insulating barrier with reduced order parameter. By increasing the particle density, we drive the system to the instability which  results in the creation of stable Josephson vortices. Finally, we show that further increase of pumping results in the destruction of the solitonic barrier and appearance of a single extended condensate, which is the ground state of the system, indicating the recovery of the superfluid behavior.

\begin{figure}[b]
  \includegraphics[width=1\linewidth]{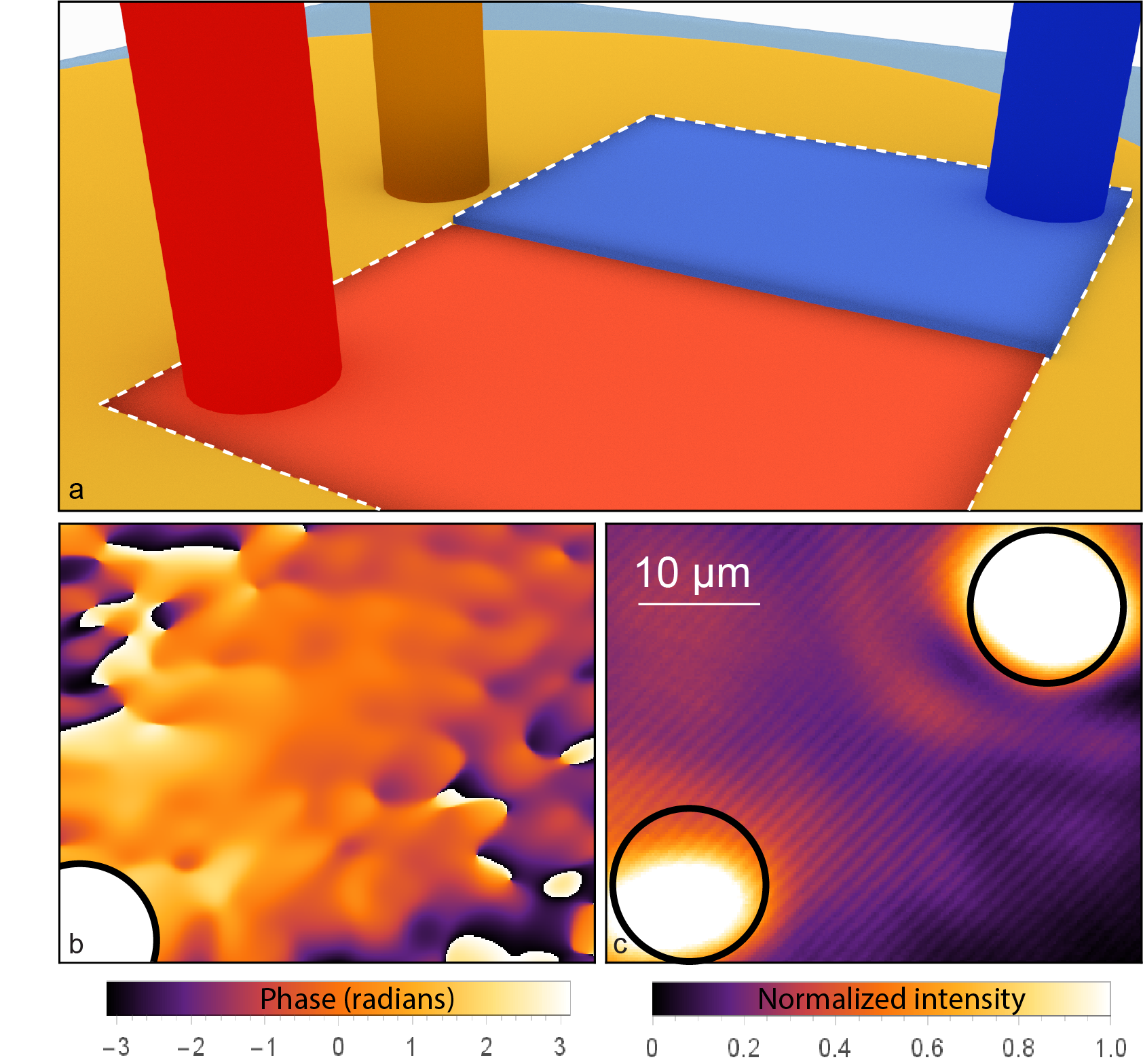} 
  \caption{
    \textbf{a, }Sketch of the polariton condensate (yellow), with the non-resonant pump shown as a dark yellow cylinder. The blue and red colors correspond to regions in which the phase of the condensate is determined by one of the resonant laser beams, denoted as blue and red cylinders, respectively. The white, dashed rectangle indicates the visible region in the measurements represented in (b, c). 
    \textbf{b, }Phase of the condensate just above the power threshold for quantum degeneracy locked to that of the external laser in the bottom-left corner of the image. Spatial scale is the same as in c). The white circle corresponds to the saturated signal of the reflected beam.
    \textbf{c, }Measured interferogram of the extended condensate. The resonant beams, partially reflected at the surface, saturate the detector at the positions of the two spots (circles) despite the small amount of injected polaritons. 
  }
\label{fig:1} 
\end{figure}
\section*{Results}
The sample used in these experiments is a high quality factor ($Q>10^{5}$) semiconductor microcavity with polariton lifetimes around $\SI{100}{\pico\second}$ (see Methods). Similar microcavities have been recently shown to exhibit ballistic propagation in the cavity plane for hundreds of microns and polariton condensation outside of the laser spot either confined by \textit{ad hoc} trapping potentials or by the same blueshifted region underneath the exciting laser ~\cite{Nelsen2013,Steger2013,Sun2017, Wertz2010,Kammann2012,Dreisman2016}.

Here, a two-dimensional polariton condensate, extending across a region much larger than the healing length, is excited by a continuous wave laser tuned well above the polariton resonances and employing a two-step relaxation mechanism as done in Ref.~\cite{Ballarini2017}. In this configuration, the dephasing induced by the high exciton density in the reservoir is limited by the ability to spatially separating the condensate from the excitation spot~\cite{Caputo2017}. The exciton reservoir is first populated through relaxation of carriers generated by the nonresonant laser, increasing the energy of the polariton dispersion (energy blueshift) in the region under the excitation spot of around $\SI{4}{\milli\electronvolt}$, see Ref.~\cite{Ballarini2017}. Polaritons at high energy are accelerated radially outward from the excitation spot and, through the joint effect of high velocity propagation and phonon assisted relaxation, they populate the bottom state of the polariton band even at distances of hundreds of micron from the blueshifted region~\cite{Anton2013,Ballarini2017}. Eventually, stimulated scattering leads to the formation of an extended polariton condensate all around the injecting spot in the ground state of the dispersion ($k\approx0$, with $k$ the polariton wavevector component parallel to the plane of the cavity)~\cite{Ballarini2017}.
In the sketch of Fig.~\ref{fig:1}a, a portion of the polariton condensate is highlighted by a dashed white rectangle to represent the region under consideration in the following measurements. The phase $\phi(\vec{r})$ of the polariton condensate is obtained in two-dimensions from the interference fringes between the signal emitted by the condensate and a reference with a spatially uniform phase (see Methods).

We show now that the phase of the polariton condensate can be locked to that of an external laser, acting as a seed in the symmetry-breaking process, tuned to resonance with the energy of the condensate~\cite{Wouters2008, Eastham2008}. 
This is shown in Fig.~\ref{fig:1}b, where the phase of the condensate is locked to that of an external laser beam, resonant with the condensate frequency and focused in the bottom-left corner of the image (see Methods and Supplementary Information). Note that the phase locking of the condensate is non-local: while the external laser is focused into a small spot (white circle in Fig.~\ref{fig:1}b) of radius $r=\SI{5}{\micro\meter}$ and it is kept at low enough power to induce only a negligible contribution to the condensate density, the phase locking extends over the whole condensate, creating a domain of uniform phase all across the region of interest (coherence lenght of 50--100~$\SI{}{\micro\meter}$, see Methods). 

As shown in the sketch of Fig.~\ref{fig:1}a and in the experimental interferogram in Fig.~\ref{fig:1}c, here we impose a twisted boundary condition to the condensate by focusing a second, resonant external beam with a different phase at a distance of about $50~\mu m$ (top-right corner in the interferogram shown in Fig.~\ref{fig:1}c). Note that the healing length is one order of magnitude smaller, $\xi\approx5~\mu m$~\cite{Caputo2017}.  
Differently from the phase imprinting schemes first used with atomic condensates \cite{Denschlag2000, Cataliotti2001}, in our configuration the phase is imposed only at the top-rigth and bottom-left corner in Fig.~\ref{fig:1}c, while it is unconstrained in the region between the two phase-locking points.  
When two concurrent beams with a phase difference are acting on the condensate, two phase domains can be expected as a result of the competing phases of the locking lasers \cite{Janot2013}. 
This arrangement is depicted in Fig.~\ref{fig:1}a, with the red and blue 
rectangles marking the two regions of uniform phase. 
We now measure the actual response of the condensate to the twisted boundary condition for different polariton densities.

\begin{figure}[h]
  \includegraphics[width=0.95\linewidth]{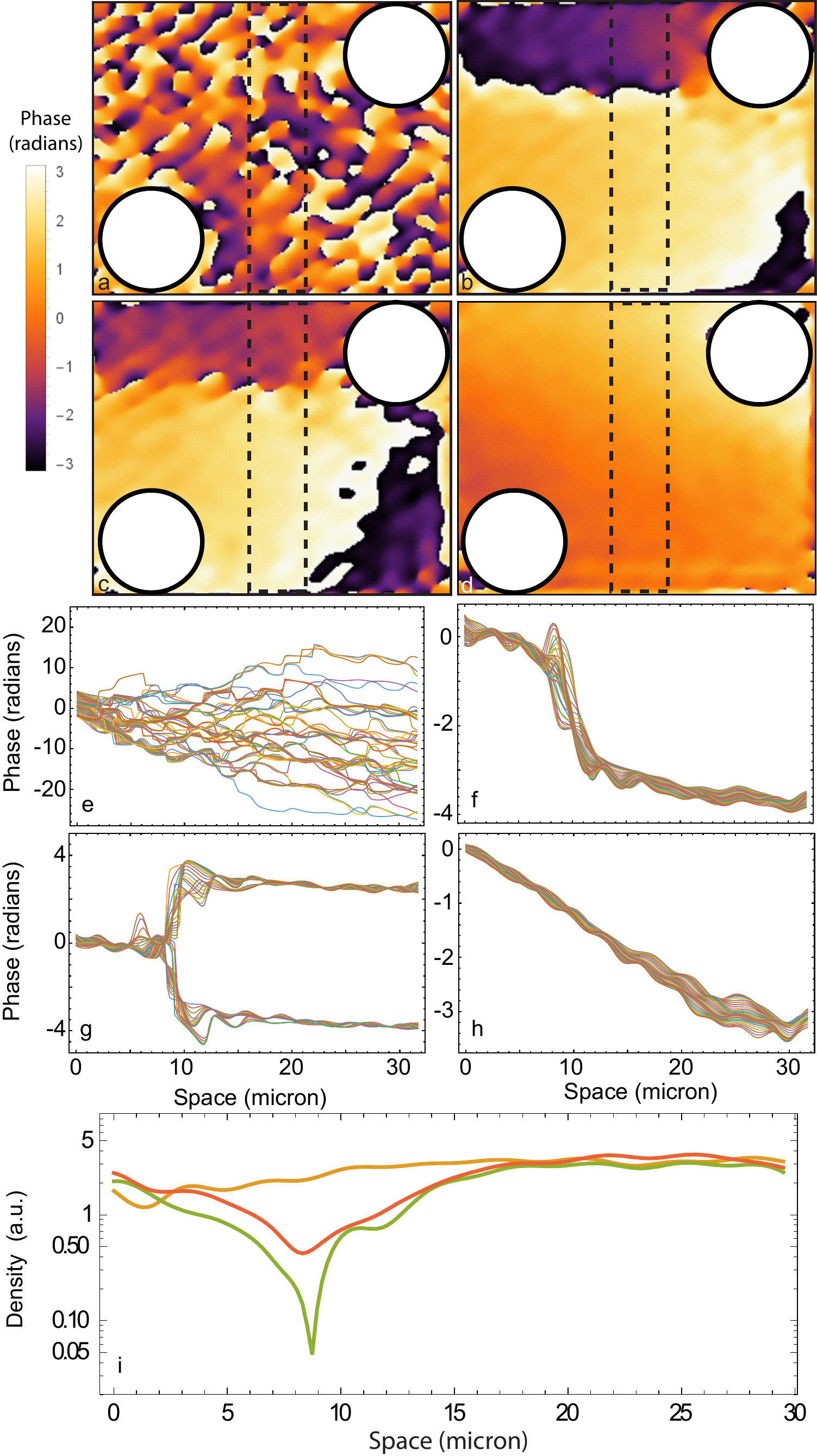} 
  \caption{   
    \textbf{a-d, }Phase (mod $=2\pi$) of the condensate with twisted boundary condition for increasing intensities of the nonresonant pump. The dashed rectangle in (a-d) corresponds to $8\times\SI{35}{\micro\metre^2}$. \textbf{a, }Below condensation threshold ($d=0.5~d_{th}$). \textbf{b, }($d=1.5~d_{th}$). \textbf{c, }($d=2.0~d_{th}$). \textbf{d, } ($d=2.5~d_{th}$).
    \textbf{e-h, }Phase profiles along the vertical direction spanning the region within the dashed-black rectangle in \textbf{a-d}, respectively. 
        \textbf{i, } Densities profiles extracted from the interferogram corresponding to the phase map shown in \textbf{b, c, d} in green, red and yellow lines, respectively.}
\label{fig:2} 
\end{figure}

The polariton density can be increased by increasing the power of the nonresonant pump, without changing the intensity of the two phase-locking beams. We note that our photoluminescence measurements are detecting the steady state of the system, averaged temporally due to the time-integrated detection (few milliseconds). Therefore, while the microscopic evolution of the system may change in different realisations depending on the possibly slightly different initial conditions, the more stable steady state solution is naturally captured in the experiments. When the polariton density is below the condensation threshold, $d_{th}\approx\SI{0.5}{pol\per\micro\metre\squared}$, a macroscopic phase is not defined in the region between the two resonant beams and the phase fluctuates from point to point (Fig.~\ref{fig:2}a). In Fig.~\ref{fig:2}e, the unwrapped phase profiles along the vertical direction in the black rectangle of Fig.~\ref{fig:2}a show the random phase oscillations in space. In Fig.~\ref{fig:2}b-d, the polariton density is brought above the condensation threshold, allowing the double locking of the condensate. For densities $d\approx1.5~d_{th}$, the hierarchy of excitations induced by the phase imprinting (phase difference of $\delta \phi \approx \pi$) ends up with a neat separation between two regions of the condensate, spatially separated by a wavy shaped junction for about $30~\mu m$ as shown in Fig.~\ref{fig:2}b. The phase profile across the junction (Fig.~\ref{fig:2}f) shows a steep phase jump of $\Delta\phi \approx \pi$, with a corresponding depletion in the density profile (green line in Fig.~\ref{fig:2}i), demonstration of the spontaneous appearance of a dark soliton-like structure acting as an insulating barrier.

In analogy to what was observed in atomic Bose-Einstein condensates~\cite{Denschlag2000}, optical nonlinear waves~\cite{Law92} and one-dimensional polaritonic wires~\cite{Goblot2016}, the formation of a soliton is an expected solution of the imposed phase boundary condition~\cite{Janot2013}. 
However, even if the topological nature of the junction could partially explain the relative robustness of this configuration against noise, it is known that, in two-dimensional systems, transversal modulation instabilities (snake instabilities) induce the transformation of dark solitons into vortex-antivortex pairs, vortex ring, vortex dipoles, or even more complex dynamics \cite{Anderson_Snaking,Ma2010,Verma2017, Gallemi2016}. 
The spontaneous formation and stability of such a topological excitation is a notable consequence of this phase imprinting scheme in driven/dissipative systems.

For increasing densities ($d \approx 2.0~d_{th}$), the solitonic junction is still present but the phase difference is inverted in some points along the nodal line (Fig.~\ref{fig:2}c). The presence of vortices and regions with an inverted current is captured by the opposite $\pi$ shift of the phase in Fig.~\ref{fig:2}g. Eventually, at the highest pumping powers employed in these experiments, the junction completely disappears, leaving a shallow phase gradient all across the condensate (Fig.~\ref{fig:2}d and Fig.~\ref{fig:2}h). At these densities ($d \approx 2.5~d_{th}$), the phase stiffness is enough to avoid phase jumps and density depletions, recovering the superfluid behavior in the region under consideration \cite{Janot2013}.

\begin{figure}[h]
  \includegraphics[width=0.95\linewidth]{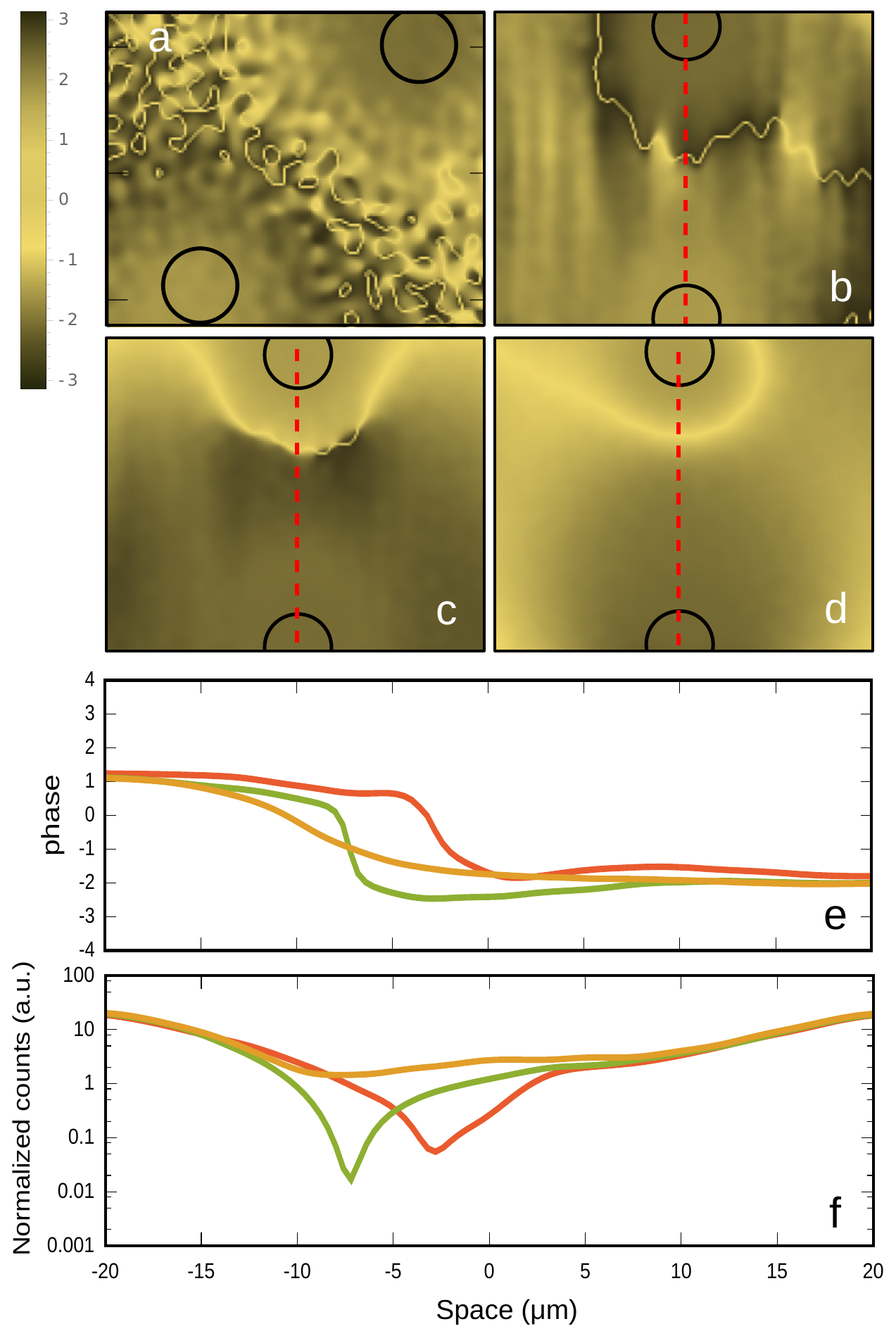} 
  \caption{    
    \textbf{a-d} Phase profiles obtained numerically with increasing pump power for {\bf a} $P\ll P_{th}$,  {\bf b} $P=4.5~P_{th}$,  {\bf c} $P=6~P_{th}$,  and {\bf d} $P=8~P_{th}$. {\bf e, f} Cross sections of {\bf e} phase and {\bf f} density corresponding to the dashed lines in {\bf b-d}. Note that in {\bf b-d} orientation of the cross section with respect to pumping spots was chosen such that the line is approximately perpendicular to the barrier. Phase is expressed in radians.}
\label{fig:2a}
\end{figure}

\begin{figure}[htbp]
  \includegraphics[width=1\linewidth]{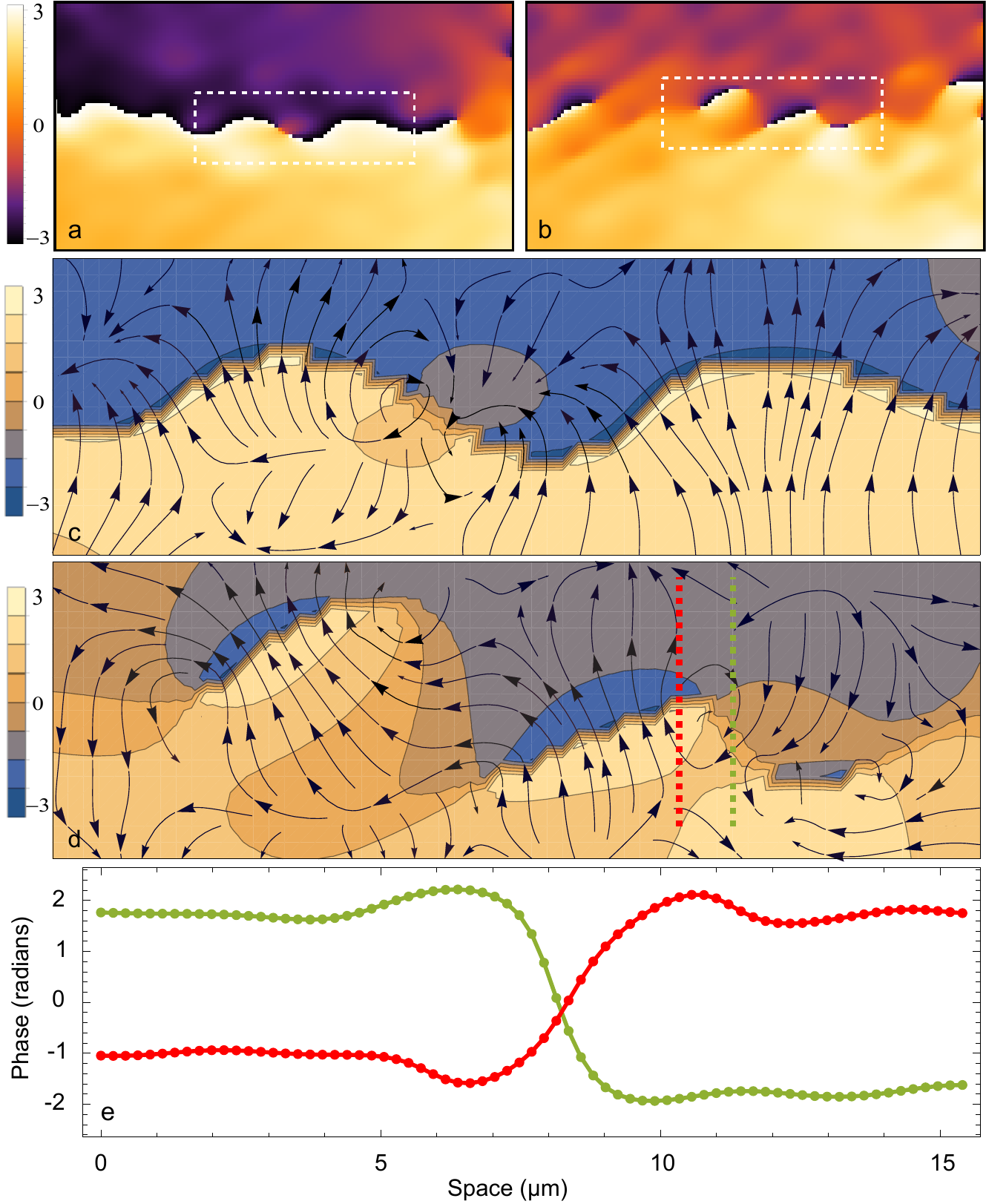} 
  \caption{
    \textbf{a, b,} Detail of the region with the phase jump in Fig.~\ref{fig:2}b and Fig.~\ref{fig:2}c, respectively, with the appearance of a wavy shaped junction.
    \textbf{c, d} Velocity streamlines directly obtained as the 2D gradient of the phase from the experimental data in the dashed-white rectangle in \textbf{a,b,} respectively. The arrows indicate the direction of the polariton currents, however the length does not indicate the intensity of the velocity field, which is decreasing away from the junction. The background images in \textbf{c, d,} are the experimental data of the phase shown in the dashed-white rectangle in \textbf{a, b,} respectively. The 2D contour plot representation is used to show the velocity streamlines orthogonal to the isophase lines. Color bar legends are in radians. In \textbf{c,} the nucleation of a vortex-antivortex pair. In \textbf{d,} the velocity streamlines show the proliferation of Josephson vortices (2-3 vortex-antivortex pairs are visible). The vortex between the dotted lines is taken as an example with clockwise rotation. \textbf{e, }Phase profile along the dashed lines in \textbf{d, }showing the inverted currents around the Josephson vortex. The whole vertical extension of $\SI{15}{\micro\meter}$ as in \textbf{a, b, }is given. The area in the white-dashed rectangle is $5\times\SI{15}{\micro\metre^2}$.
  }
\label{fig:3} 
\end{figure}

We reproduce numerically the above stages of formation and disappearance of coherent structures using the complex Ginzburg-Landau equation, adapted to the case of an exciton-polariton condensate separated from the exciton reservoir~\cite{Caputo2017} (see Methods). 
In Figs.~\ref{fig:2a}a-d we show the results of simulations carried out with increasing pump power, qualitatively reproducing the formation of a solitonic barrier, the nucleation of vortices and finally the appearance of a homogeneous condensate. The figures show snapshots of the phase of the wavefunction after long time of evolution, when the initial transient effects have washed out, and the system has approached the steady state.  In Figs.~\ref{fig:2a}e-f, cross sections of phase and density profiles are shown. We find that formation of the barrier and vortices observed experimentally is reproduced in simulations. 
The phase jumps imprinted by the pump together with the noise present in the stochastic equation act as a seed for the appearance of solitons and vortices. In the experiment, the noise is not stochastic but rather given by very small local fluctuation of the photonic or exciton potential~\cite{PhysRevLett.106.115301}. 
In the absence of disorder potential in the simulation, the positions of solitons and vortices are fluctuating. However, we find numerically that a weak static disorder is able to pin the topological excitations in certain fixed positions, without altering their physical structure, exactly as it happens in the experiment. 
At the same time, the potential disorder required for pinning is very weak and by itself does not form an insulating barrier. Phase defects appear only when the double-locking is applied, leaving otherwise a flat phase over the whole region (see Supplementary Information).
Since we do not assume any particular form of disorder in the simulation, the soliton is not pinned to the same point in the sample as in the experiment, but rather its position is determined by a particular realization of the noise and disorder, which vary from one simulation to another.

In Fig. \ref{fig:3}, we focus on the region in proximity to the junction for the intermediate power regime. The phase shown in Fig.~\ref{fig:2}b and Fig.~\ref{fig:2}c is magnified in Fig.~\ref{fig:3}a and Fig.~\ref{fig:3}b, respectively. The velocity vector field can be obtained directly from the measured phase $\phi(\vec{r})$ by $\vec{v}(\vec{r})\propto\nabla\phi(\vec{r})$~\cite{Nardin2011,Dominici2015,Gianfrate2018}. In Fig.~\ref{fig:3}c and Fig.~\ref{fig:3}d, the velocity streamlines corresponding to the data in the white-dashed rectangle in Fig.~\ref{fig:3}a and Fig.~\ref{fig:3}b, respectively, are shown. 
In analogy to the formation of Josephson vortices in superconducting Josephson junctions or in atomic Bose-Josephson junctions \cite{Kaurov2005, Roditchev2015}, the nucleation of vortices in the barrier allows a partial dissipation of the energy stored in the twisted phase configuration. This is markedly different from previous works on solitons and vortices in polariton superfluids, where hydrodynamic effects across the condensate prevails over the dynamics of Josephson vortices within the barrier~\cite{Tosi2012,Hivet2014,Ohadi2016}. 
The appearance of the first couple of Josephson vortices is evident in Fig.~\ref{fig:3}c, where a vortex-antivortex pair (formed by vortices with opposite circulation) nucleates within the barrier and supports an inverted current going downward (center of Fig.~\ref{fig:3}c), as opposed to the upward flux seen across the rest of the junction.
 
At higher densities, the separation between the vortex and the antivortex increases and the nodal line breaks also at different points, as shown in Fig.~\ref{fig:3}d, developing polariton currents in opposite directions across the junction. The solitonic features in the density and phase profiles are still visible, but now the nodal line has a more complex structure (Fig.~\ref{fig:3}d). 
As shown in Fig.~\ref{fig:3}e, the phase slip is inverted along the red and green dashed-lines in Fig.~\ref{fig:3}d, going from $\Delta\phi \approx \pi$ to $\Delta\phi \approx -\pi$ in accordance to the inverted currents in nearby domains of the junction. 
Moreover, transversal currents induced by the vorticity are visible in the condensate at both sides of the junction, that now acquires a more tortuous path, with additional phase modulations departing from the vortex cores at finite angles from the main nodal line. At higher densities, all these topological excitations disappear and the system is left in a clean superfluid phase with a smooth phase gradient between the boundaries.
\section*{Discussion}
The phenomenology of the transition from a solitonic structure to vortex pairs may appear similar to transverse (snake) instability of dark solitons observed in several nonlinear two-dimensional systems~\cite{Mamaev_Snaking,Tikhonenko_Snaking,Anderson_Snaking, Frantzeskakis_Darksolitons}. 
To the contrary, we emphasize that our observations are not an example of such instability, since the observed structures are stable at all values of pumping power. Snake instabilities in conservative systems develop in time, with the evolution starting from an elongated dark soliton state that is prepared intentionally, e.g., by a careful phase imprinting. The fragility of a soliton solution of the nonlinear Schr\"odinger equation results in unavoidable decay into vortex-antivortex pairs, even when phase difference is imposed as a boundary condition. The vortices resulting from snaking instability follow a dynamical evolution, move away from the initial position, which results in a complete decay of solitonic structure after certain amount of time. In contrast, in our pumped dissipative system, resonant lasers provide effective phase locking which results in stable two-dimensional solitonic structure. At higher pump power, the soliton splits into vortex pairs, which however stay in positions close to the initial soliton dip, and the insulating character of the barrier is partially retained as follows from the phase profile in Fig.~\ref{fig:2}g and Fig.~\ref{fig:3}e. Moreover, the last stage of the transition shown in Fig.~\ref{fig:2}d, with a smooth phase and flat density, is never observed as a result of instability in a conservative system. This can be understood in the following way.
The increase of pumping drives our non-equilibrium condensate to lower energy states as a result of faster thermalization rate~\cite{Yamamoto_PiState}. This leads to the transition from the soliton-like state, through proliferation of vortices, to the flat density state which is closest to the ground state of the system (see Methods for detailed calculation of energy of particular states). Hence, the transition from the solitonic state to vortices to a flat state is a result of gradual relaxation transition between stable steady states, rather than an instability of an artificially prepared out-of-equilibrium state which decays unavoidably. From the point of view of potential applications, the stability of these steady states can be a significant advantage as compared to unstable states.

In conclusion, we have shown the full range of dynamical response of a polariton condensate to a twisted-phase boundary condition by changing the polariton density and looking at the steady state evolution of the system.
The existence of a stable spatial soliton with a wiggling nodal line, as in Fig.~\ref{fig:2}b, is a remarkable topological feature and constitutes a natural realisation of a bosonic LJJ in polariton condensates~\cite{Reinhardt1997}. The solitonic structure is characterized by a strong reduction of the order parameter (condensate wavefunction amplitude), which acts as an analog of the normal barrier, or weak link, in superconducting Josephson junctions, and separates two extended regions of the condensate with well defined phases.
Remarkably, the transition from the superfluid behavior to the formation of stable topological excitations can be controlled in the same experiment by optical tuning of the polariton density, suggesting the possibility to test the scaling laws of interacting quantum fluids~\cite{Su2013} in a solid state environment.

\begin{acknowledgements} DC, DB and DS acknowledge the ERC project POLAFLOW (Grant N. 308136) and the ERC "ElecOpteR" grant N. 780757. NB and MM acknowlegde support from National Science Center grants 2015/17/B/ST3/02273 and 2016/22/E/ST3/00045. Enlightening discussions with Marzena~H.~Szyma\'nska are acknowledged.\end{acknowledgements}

\section{Methods}

\subsection{Sample and experimental setup}

The sample is a high quality-factor $3/2 \lambda$ $GaAs/(Al,Ga)As$ 
two dimensional planar cavity with 12 GaAs quantum wells placed at three antinode 
positions of the electric field. The top (bottom) layer of mirror (Distributed Bragg Reflectors, DBR) includes $34$ ($40$) 
pairs of $AlAs/Al_{0.2}Ga_{0.8}As$ layers. The effective Rabi splitting of the sample is
 $\SI{16}{\milli \electronvolt}$. The cavity-exciton energy detuning dependes on the 
spatial position, consequently we used a point with a slightly negative value,
$\delta=\SI{-2}{\milli \electronvolt}$.
Experiments are performed using a nonresonant continuous wave excitation with a low-noise, 
narrow-linewidth Ti:sapphire laser with stabilized output frequency (M Squared Lasers) and chopped with a 4 kHz frequency and duty cycle around 5\% to avoid heating of the sample.
In addition to the non-resonant pumping, a resonant laser, splitted into two beams and with momentum and energy matching the bottom of the lower polariton branch, is employed to pin the phase of the condensate. The two resonant beams are produced in a compact interferometer and belong to the same continuous wave laser, maintaining therefore the same phase difference. The stability of the interferometer is optimised to minimize the residual uncertainty in the phase difference between the two beams, which is adjusted to be $\approx\pi$ radians and can be controlled by tuning the position of the mirror in one of the two arms of the interferometer. The formation of the phase slip is robust enough to be observed also in the presence of the unavoidable fluctuations of the phase difference between the two resonant beams during the time of the experiment, without needing any active feedback for further stabilization of the phase difference. However, the stability of the topological structures quickly deteriorates when the nominal phase difference is shifted from the $\pi$ radians value.
The wavelength and the angle of incidence of the resonant beams are adjusted very carefully in order to match the resonance, both in energy and wavevector, of the polariton condensate. This is done by recreating the Fourier plane of the launching objective in a 4-f configuration, in order to control the component of the wavevector parallel to the sample surface with a precision of $\SI{0.1}{\micro\meter^{-1}}$.

The sample emission is collected and sent to a Mach-Zehnder interferometer and finally focused on the entrance slit of a CCD camera.
The polariton lifetime is determined by time-resolved photoluminescence measurements with a Ti:sapphire laser delivering 3~ps pulses with repetition rate of 82~MHz. Polaritons are resonantly injected by tuning the frequency and angle of incidence of the beam to match the polariton resonance close to the bottom energy of the dispersion. Resonant excitation is required to prevent the radiative decay rate measurements from being affected by the bottleneck effect and the long lifetime of the excitonic reservoir. The photoluminescence signal is recorded with a streak-camera with an overall time resolution of 3~ps and the mono-exponential decay easily allows to extract a lifetime of 100~ps~\cite{Steger2013,Ballarini2017}.

\subsection{Phase measurements}

Measurement of the phase of the condensate are performed with a Mach-Zehnder interferometer, where the signal emitted by the microcavity is interfered with a reference beam of spatially uniform phase, obtained by expanding a single spatial point of the condensate, chosen far away from the region of interest.
The phase difference from the reference value is extracted at each position of the condensate from the pattern of the interference fringes by using the standard Fast Fourier Transform algorithm~\cite{Caputo2017}. 
To demonstrate the non-local locking of the phase of the condensate by the external laser, we use the Mach-Zehnder interferometer to interfere a region of the condensate outside of the laser spot with a reference signal taken from the same laser used to pin the phase. 
In our configuration, the coherence of the condensate is not significantly altered by the phase locking with an external laser, leaving unaffected the coherence length that, as shown in \cite{Caputo2017}, depends only on the polariton density of the condensed state. In particular for our measurement of long Josephson junctions we have a power law decay of coherence with an exponent of $\alpha=0.5$ which corresponds to a coherence length of $L_{c}=\SI{50}{\micro\meter}$, while for the case of the superfluid regime the exponent is $\alpha=0.25$ with a coherence length $L_{c}>\SI{100}{\micro\meter}$.

\subsection{Numerical modeling}

We model the polariton system using the stochastic complex Ginzburg-Landau equation~\cite{Caputo2017}
\begin{eqnarray}
\label{SGPE}
i\frac{\textrm{d}\psi}{\textrm{d}t}&=&\Big[-\frac{\hbar}{2m^*}\nabla^2+g|\psi|^2+ V({\bf r})
+i\left(\gamma - \kappa - \Gamma|\psi|^2\right)\Big]\psi +\nonumber\\ &+& F_1({\bf r}) {\rm e}^{-i \omega_L t} + F_2({\bf r}) {\rm e}^{-i \omega_L t+i\Delta\phi}
+\frac{\textrm{d}W}{\textrm{d}t},
\end{eqnarray}
where $\psi({\bf r},t)$ is the complex wave function, $m^*$ is the effective mass of lower polaritons, $g$ is the interaction coefficient, 
$V$ is the weak disorder of the sample, $\gamma$ is the effective pumping rate,
$\kappa$ is the polariton loss rate, and $\Gamma$ is the saturation coefficient. The spatial functions $F_1$ and $F_2$ describe the Gaussian profiles of the two resonant pumping lasers, $\omega_L$ is the laser frequency and $\Delta\phi$ the phase difference. The strength of the disorder $V({\bf r})$ required for spatial pinning is chosen to be  lower than other typical energy scales in the system, and its spatial correlation length is a few micrometers. 
The term $dW$ is a scaled Wiener noise with correlations
\begin{align}
\label{noise-phi}
\langle\textrm{d}W({\bf r},t)\textrm{d}W^*({\bf r}^{\prime},t)\rangle&=c_{\rm q}\frac{\gamma + \kappa + \Gamma|\psi|^2}{dV}\delta_{\bf r,r'}\textrm{d}t,\nonumber
\end{align}
where $c_{\rm q}$ is a parameter representing the relative strength of quantum fluctuations present in the system.
Note that since our condensate is spatially separated from the hot reservoir created by the pumping laser, we do not include a separate equation for the dynamics of the reservoir density.

The above form of the stochastic equation has been derived using both the perturbative Keldysh field theory~\cite{Diehl_KeldyshReview} and the truncated Wigner approximation~\cite{QuantumFluids}. We note that as we are not investigating the region close to the phase transition, we can neglect the correction to the wavefunction due to the ordering of operators~\cite{Caputo2017}. The values of parameters extracted from the experimentally measured dispersion are $m^*=3.85\times10^{-5}m_{\rm e}$ and $\kappa=(100\, {\rm ps})^{-1}$. The interaction coefficient is $\hbar g=4\times 10^{-3} {\rm meV} \mu {\rm m}^2$, the saturable nonlinearity $\hbar \Gamma = 14 \times 10^{-3} {\rm meV} \mu {\rm m}^2$, and the noise coefficient $c_{\rm q}=0.1$. The blueshift of the condensate at threshold is $\hbar g d_{th}=\SI{0.05}{\milli \electronvolt}$, the frequency of the resonant lasers $\omega_L$ is set to the lower polariton energy at threshold, and their spatial FWHM is 10$\mu$m.

\subsection{Energy of the condensate}

Consider the energy of the condensate in the state appearing at the highest pump power, as in Fig.~\ref{fig:2}d. Energy is composed of the kinetic and potential parts. The kinetic part is associated with the phase gradient between the pinning laser spots, and the potential energy is due to repulsive polariton interaction
\begin{equation*}
  E_{\rm 0}=  \int \left[\frac{\hbar^2}{2m^*}|\nabla \psi|^2  + \frac{\hbar g}{2} |\psi|^4 \right] d \er \approx (A / d^2 + B)S, 
\end{equation*}
where $S$ is the approximate surface of the condensate, $d$ is the distance between the phase pinning beams, and $A$ and $B$ are constants.

The energy of the dark soliton-like state (Fig.~\ref{fig:2}b) is 

\begin{align*}
  E_{\rm ds} &= \int_{\rm soliton} E\, d\er + \int_{\rm rest} E \,d\er \approx \\ &\approx (A' / w^2 - B') S_{\rm soliton} + B (S - S_{\rm soliton})= \\
    &=(A' / w - B' w) L + B (S - w L),
\end{align*}
where we integrate separately over the extent of the soliton and the rest, $L$ is the length of the soliton, $w$ is its transverse width, and $A'$ and $B'$ are constants corresponding to kinetic and interaction energy in the soliton region. In the limit of thin and long dark soliton, with $x\ll d, L$ and $wL \ll S$, the dominant contribution in the difference between $E_{\rm ds}$ and $E_0$ is $E_{\rm ds}- E_0\approx A'L/w$, which is positive. Consequently, the dark soliton state is an excited state of the system, and the state with a small constant phase gradient has a lower energy.

Data availability: The raw experimental and numerical data used in this study are available from the corresponding author upon reasonable request.

%
\bibliographystyle{apsrev4-1}

\end{document}